\RequirePackage{fix-cm}
\documentclass[smallextended]{svjour3}       % onecolumn (second format)

\usepackage[latin9]{inputenc}
\usepackage{color}
\usepackage{url}
\usepackage[authoryear]{natbib}

\smartqed  % flush right qed marks, e.g. at end of proof
\usepackage{graphicx}
\def\makeheadbox{\relax}

\begin{document}

\title{Likelihood criteria for the universe%\thanks{Grants or other notes
}
% \subtitle{Do you have a subtitle?\\ If so, write it here}


\author{Ezequiel L\'opez-Rubio}


\institute{E. L\'opez-Rubio \at
              Universidad de M\'alaga \\
              Bulevar Louis Pasteur, 35, 29071 M\'alaga, Spain\\
              Tel.: +34-952-137155\\
              Fax: +34-952-131397\\
              \email{ezeqlr@lcc.uma.es}           %  \\
}

\date{}

\maketitle

\begin{abstract}
The development of science and technology has progressively demonstrated the ability of humankind to understand and manipulate the physical world, and it has also shown some fundamental limitations to predictability of physical events. This realization has led many thinkers to wonder why the universe has the observed level of regularity. Justifications of this fact tend to present our universe as a likely option among some range of possibilities. In this work, an assessment of the likelihood criteria employed for such justifications is carried out. Furthermore, four alternative universes are described that appear to be more likely than our own, depending on the likelihood criterion that is considered.
\end{abstract}

\section{Introduction}

The comprehensibility of the universe has long been recognized as an intriguing mystery that stands as one of the fundamental issues in the philosophy of physics\footnote{`One may say ``the eternal mystery of the world is its comprehensibility'' ' (\citealp{Einstein:1936}, p. 351). }. Scientists and philosophers of all times have noticed that the highly regular structure of reality is a nonobvious fact. Wigner remarked that the effectiveness of mathematics to describe the physical world is deeply surprising (\citealp{Wigner:1960}). This situation poses the question of whether physical laws have an existence on their own. Still, the current state of matters in quantum mechanics indicates that there are essentially unpredictable events at the microscopic scale.

Physics is based on the principle that the universe is intelligible, and this principle is not self-evident (\citealp{Davies:2008}, chapter 1; \citealp{Livio:2010}, chapter 9; \citealp{Coyne:2008}, pp. 28-29). In other words, it is imaginable that the universe could be less amenable for rational inquiry. The multiverse explanation to the fine tuning of physical constants neglects the possibility of universes with fundamentally different physical laws (\citealp{Ratzsch:2005}, p. 674). Therefore, the range of possible options considered by the multiverse approach is restricted to those allowed by the multiverse version of our physical theories. However, the universe (or the multiverse) could have had other laws or none at all (\citealp{Ellis:2007}, p. 1241; \citealp{Susskind:2006}, Introduction). Furthermore, it could have had different laws depending on the region (\citealp{Davies:2008}, chapter 8). Little effort has been devoted to studying alternative physical laws since it is difficult to establish the set of all possible laws (\citealp{Barnes:2012}, p. 536). It is hard to generate a set of alternative physical laws that completely depart from ours and then figure out how such laws could give rise to intelligent life.

This work is devoted to the analysis of the criteria to evaluate the likelihood of our universe. Those positions are closely related to the perceived regularity of our universe, as compared to other imaginable universes. A quantitative approach is chosen for this investigation so that the likelihood of the different options can be compared. Four alternative universes are detailed, which pose significant difficulties to justify the likelihood of our universe.

\section{Preliminaries\label{sec:Preliminaries}}

In this work, some concepts will be used with a quite specific meaning. Next, some definitions are given of these concepts that will be employed later. 

%MAYBE THE COMPREHENSIBILITY DEFINITION IS NOT NECESSARY. A description of the universe is regarded as comprehensible if it can be understood by a contemporary human with proper training. Comprehensibility is shaped by evolution, culture, and history so that the ability of humans to grasp descriptions of physical reality has progressed across the epochs. Descriptions that are understood by more contemporary people are considered simpler than others. In case that intelligent beings exist outside our planet, they could have other cognitive abilities. Hence their comprehensibility criteria would probably be different than ours.

A history of a region of spacetime is defined as a description of the state of the space comprised by the region at successive time instants within the time interval of the region (\citealp{Knobe:2006}, p. 49). In other words, a history is a sequence of states of the contents of the region through time. If quantum mechanics is assumed to hold, then for each finite region of space, there is a finite amount of possible states, as stated by the holographic entropy bound (\citealp{Bekenstein:1981}, p. 287; \citealp{Susskind:1995}, p. 6377). Therefore, the state of each finite region of space can be completely described by a finite amount of information. In this context, a history is a sequence of complete descriptions of the state of the region at successive Planck time intervals. All histories that do not violate exact conservation laws have a nonzero probability of occurrence (\citealp{Knobe:2006}, p. 49). Quantum mechanics assigns a specific probability to each history. This is equivalent to saying that quantum mechanics assigns a transition probability from each state to each state. In what follows, quantum mechanics is not assumed to hold, but I will still assume that for any region of spacetime, there is a finite number of possible histories of the region so that each history can be characterized by means of a finite amount of information. 

In this paper, the term `anarchic' has a specific meaning. It does not refer to chaotic systems that evolve according to the known laws of physics. In this paper, I say that a region of spacetime of a universe is anarchic whenever the state transitions in that region do not conform to the known laws of physics. For example, it could be that all state transitions happen with the same probability, which would not agree with quantum mechanics. In other words, `anarchic' means that it does not obey the known laws of physics. Hence `anarchic' is fundamentally different from the `chaotic' term employed to pose the possibility of Boltzmann brains in a universe that follows our laws of physics. Such brains arise as thermal fluctuations surrounded by a disordered, high entropy environment that follows the known laws of physics (\citealp{davenport2010boltzmann}, p. 1; \citealp{NOMURA2015514}, p. 518). Under some circumstances, Boltzmann brains turn out to be much more likely than human-like intelligent observers (\citealp{davenport2010boltzmann}, p. 8; \citealp{Barnes:2012}, p. 561) because the former require a smaller degree of organization. Furthermore, it has been argued that Boltzmann brains emerge even in universes not fine tuned for life (\citealp{Collins:2012}, pp. 175-176). 

I will consider the concept of a string in the automata theory sense. This means that, given a finite set of symbols called the alphabet, a string is a finite sequence of symbols from the alphabet. It is important to note that for any given alphabet, there is a countable infinity of possible strings on that alphabet.

The description $D\left(x\right)$ of a universe $x$ is a string made of three components, $D\left(x\right)=d_{1}\left(x\right)d_{2}\left(x\right)d_{3}\left(x\right)$, which are the minimum necessary to specify what happens anywhere at any time in such universe:
\begin{enumerate}
    \item A description of the initial conditions of the universe, noted $d_{1}\left(x\right)$.
    \item A mathematical specification of the laws of the universe, if it has any, noted $d_{2}\left(x\right)$.
    \item A specification of all events that are not logically entailed by the previous two components, noted $d_{3}\left(x\right)$.
\end{enumerate}
As known, quantum mechanics implies that there are many events in our universe that cannot be predicted; for example, in a double-slit experiment. Therefore, for our universe $x_{ours}$ the specification of all unpredictable events associated with quantum indeterminacy is included in the third component of its description $D\left(x_{ours}\right)$.

It is worth noting that all descriptions $D\left(x\right)$ are finite because they are strings. This applies to finite universes and infinite universes that can be described finitely, such as a universe that contains infinitely many copies of some region. For the purposes of this discussion, we may restrict our attention to the observable part of the universe, which is finite. Since descriptions are finite, no numeric value in them can be specified with infinite precision. This is, in fact, advantageous since infinite precision numbers are idealizations that render the laws of physics uncomputable and hence useless for any predictive purpose (\citealp{Davies:2008}, chapter 10). Please note that it is still possible to specify numbers with an arbitrarily high (but not infinite) degree of precision by using arbitrarily large (but finite) strings. Therefore, the precision in the specification of any physical constants or magnitudes could in principle be as exact as required. Another consequence of the descriptions being finite is that the number of possible universes is a countable infinity since there cannot be more descriptions than strings.

Descriptions of the universe vary in their simplicity. Simplicity depends on the person that analyzes the description. Let us note $S\left(D\left(x\right)\right)$ the simplicity of a description $D\left(x\right)$ of a universe $x$, where it is assumed that simplicity is quantified as a number: the simpler the description $D\left(x\right)$, the larger the simplicity $S\left(D\left(x\right)\right)$. Of course, one could define several simplicity criteria, which would lead to several functions $S_1$, $S_2$,... For the sake of our discussion, we will consider just one simplicity criterion $S$ because our argumentation does not depend on the details of the simplicity criterion.

The length of the additional specification of unpredicted events $d_{3}\left(x\right)$ depends on how detailed the predictions of the laws $d_{2}\left(x\right)$ are for a set of initial conditions $d_{1}\left(x\right)$. The more predictive the laws, the shorter the third component of the description of a universe, which leads to a larger simplicity $S\left(D\left(x\right)\right)$. There are two extreme possibilities in terms of predictability:

\begin{itemize}
    \item A deterministic universe which is fully described by a set of laws has an empty specification of unpredicted events $d_{3}\left(x\right)$.
    \item An anarchic universe which does not exhibit any regularity has an empty set of laws $d_{2}\left(x\right)$.
\end{itemize}

Our universe is not completely anarchic, as demonstrated by our success in building tools and machines. It might be completely deterministic by some set of yet unknown physical laws.

\section{Positivists versus transcendentalists\label{sec:Likelihood-criteria}}

The differences in assessing the likelihood of different kinds of universes can be traced to their underlying assumptions. Barrow (\citealp{Barrow:2002}, pp. 280-281) argues that the histories where there is lawful order and then anarchy are less likely because they need more specification than the histories where the laws of physics are always valid. On the contrary, Holder (\citealp{Holder:2004}, pp. 125-126) argues that the histories with orderly law and then anarchy are more likely because there are more histories of this kind. Please note that both Barrow and Holder refer to universes that do not conform to our physics laws because they consider situations where the universe is exactly like ours up to a certain time instant and then dissolves immediately into an utterly disordered state. Such situations are extremely improbable under our laws of physics. Hence they are not talking about chaos in the ordinary physical sense, but about anarchy in the sense specified in Section \ref{sec:Preliminaries}.

As seen, their evaluation of the likelihood of the histories of the universe with order and then anarchy is precisely the opposite. This can be explained by the difference between their criteria to assess likelihood. Barrow assumes that lawful order is more likely than anarchy, which implies that simple descriptions (with less specifications) are more likely than complex descriptions. On the other hand, Holder takes for granted the state space (also called phase space) of our universe and then makes calculations about the number of histories, which leads him to conclude that more anarchic histories are more likely because there are more anarchic stories than ordered ones. It must be noted that the state space of the universe is given by our physical laws, so with other physical laws, the state space would be different. 

No matter how complex the observable universe is, a set of physical laws can always be found that fit it, although they might be ugly (\citealp{Davies:2008}, chapter 9). So universes with an anarchic part are still explainable by a set of physical laws, but these laws are not simple. Alternatively, a universe with one or more anarchic parts could also be described by different sets of physical laws depending on the part of the universe (\citealp{Hamlin:2017}, p. 584), also leading to low simplicity of the overall description of such a universe, which would be composed of several sets of physical laws along with a specification of the part of the universe where each set of laws holds.

Barrow takes the laws of physics as the fundamental ontological category, while Holder considers that the fundamental category is that of the state space of the universe. Tegmark's mathematical universe is the extreme version of Barrow's position. Barrow and Tegmark are, therefore, Platonist. Holder's approach is anti-Platonist like Wheeler since physical laws are seen as observed regularities in Nature and not transcendent truths (\citealp{Davies:2008}, chapter 10). This split has also been termed as a division between two groups: the transcendentalists and the positivists (\citealp{Russ:2011}, p. 210). For a transcendentalist, physical laws exist in a mathematical, Platonic realm beyond the observable universe (\citealp{Davies:2008}, chapter 1; \citealp{Ellis:2017}, p. 9). The universe evolves according to those laws. On the contrary, for a positivist, only the universe exists while physical laws are just contingent descriptions of the observed regularities.

The likelihood criteria advocated by the two groups are incommensurable. Moreover, they are not precise since they are not based on a specific quantitative measure. Next, I present a proposal of operationalizations for each position, and I examine the consequences of such proposals (Sections \ref{sec:Transcendentalist-universe-selection} and \ref{sec:Positivist-universe-selection}).

\section{Transcendentalist universe selection\label{sec:Transcendentalist-universe-selection}}

Conciseness is regarded by some scientists as a criterion based on beauty, elegance, or aesthetics, to favor one physical theory over another, which is known as the ontological Occam's razor (\citealp{Riesch:2010}, p. 80; \citealp{Susskind:2006}, Epilogue). From a Platonist stance, one might say that our universe and its associated physical laws have a high likelihood because such laws are concise. Some Platonists even argue that it can be assumed that the universe is computable because the computability hypothesis is simpler than its opposite (\citealp{Schmidhuber:2012}, p. 381). Transcendentalists think that the universe follows a set of simple and comprehensible laws that exist by themselves in a mathematical realm. Therefore, they focus on the second component of the description of a universe, as defined in Section \ref{sec:Preliminaries}. However, in order to adequately assess the simplicity of a description of the universe, the three components (initial conditions, laws of physics, and unpredicted events) must be considered together. Let us justify this necessity. A possible description of the universe can be elaborated that contains just one dummy law of physics, which says that no events can be predicted so that all events are specified in the third component of the description. Such a description has the simplest possible set of laws and the most complex set of unpredicted events. It is a bare enumeration of ground facts. It would be regarded by a transcendentalist as a useless description of the universe since no predictions are made (\citealp{Coyne:2008}, p. 103). Moreover, the transcendentalist would complain that it is unlikely since it conveys no comprehensible structure for the universe, i.e., it is entirely anarchic. Therefore, the simplicity of the physical laws cannot be the sole criterion for a transcendentalist to evaluate a description of the universe. Hence the overall simplicity of the three components must be evaluated.

Here I propose the simplicity of a description of a universe as a transcendentalist likelihood criterion. Using description simplicity as a likelihood criterion is similar to postulating that many state changes of the universe must be entailed by the physical laws so that the amount of unpredictable events is kept to a minimum. Therefore Barrow's argument that order followed by chaos is unlikely amounts to saying that if we assume that the universe must follow a set of laws that are simple, then our universe has a high likelihood with respect to universes that are less simple to describe.

The above considered simplicity criterion can be expressed in precise terms. Let us note $P\left(x\right)$ the probability (likelihood) of a universe $x$. Then the criterion states that universes with higher simplicity have higher likelihood:

\textbf{Descriptive parsimony principle} (DPP): For all possible universes
$x$ and $y$, if $S\left(D\left(x\right)\right)\ge S\left(D\left(y\right)\right)$,
then $P\left(x\right)\ge P\left(y\right)$.

The probability distribution on the possible laws of physics is a fundamental issue (\citealp{Ellis:2012}, p. 140). It has been stated that a uniform likelihood criterion on the set of all possible laws of physics leads to a low likelihood of a universe like ours since intricate sets of laws are more abundant than regular ones (\citealp{Livio:2010}, chapter 9; \citealp{Page:2012}, p. 117). In fact, such a uniform criterion does not lead to a proper probability distribution on the set of possible universes, which form a countable infinity. This is because there is no way to define a uniform probability distribution on an infinite set. Hence even if we choose the laws of physics as the fundamental ontological category, this does not lead to a universe with a simple set of laws unless we add the assumption that the simplicity criterion determines the likelihood of a set of laws of physics through the DPP. Furthermore, it must be ensured that the countably infinite sum of the probabilities $P\left(x\right)$ of all universes is one. 

The number of possible sets of universe descriptions for a given description length $k$ tends to infinity as $k$ grows. In particular, it is $O(n^k)$, where $n$ is the number of symbols of the alphabet, due to the number of possible strings of length $k$ over an alphabet of $n$ symbols. If we assume the DPP, given that the growth of the number of possible sets of descriptions is exponential in $k$, it follows that the average probability $m\left(k\right)$ of all universes with descriptions of length $k$ decreases exponentially with $k$, so that we can obtain a proper probability distribution $P$. In mathematical terms, if we note:

\[
m\left(k\right)=E\left[P\left(x\right) \;|\; L\left(D\left(x\right)\right)=k\right]
\]
where $L$ denotes the length of a string, then we have that $m\left(k\right)$ decreases exponentially with $k$, where $E$ stands for the mathematical expectation operator. In particular, $m\left(k\right)=\Omega\left(n^{-k}\right)$.

 The transcendentalist position of founding its likelihood criterion on the simplicity of the description of the universe has two critical shortcomings that are discussed next.
 
Under the current standard view of quantum mechanics, elementary particles such as electrons or photons sometimes act in an acausal way, which is known as irreducible randomness, also called ontic indeterminism (\citealp{Khrennikov:2019}, pp. 1-2; \citealp{Khrennikov:2016}, pp. 207-208). In other words, their behavior is inherently unpredictable (\citealp{Martinez:2018}, p. 2). For our purposes, this means that the random events that involve them must be specified in the third component $d_{3}\left(x_{ours}\right)$ of the description of our universe $x_{ours}$ since they cannot be predicted by the physical laws. Let us consider perturbed versions of our universe, which consist of changing the quantum random fluctuations on the perturbed regions by pseudorandom fluctuations, i.e., fluctuations that seem random but can be generated by a small computer program (\citealp{Shen:2017}, p. 461; \citealp{Schmidhuber:2012}, p. 388; \citealp{Wolfram:2002}, pp. 300-301). Such pseudorandom fluctuations can be described in an extremely simple way when compared to the genuinely random fluctuations of our universe. Therefore, the perturbed versions of our universe have simpler descriptions than our universe. Hence, under the descriptive parsimony principle, the perturbed versions are more likely than our universe. Moreover, the perturbed versions would be equally hospitable to human life as our universe.
 
 Secondly, it has been found that the weak interaction of the standard model of physics is not required for intelligent life. That is, universes without weak interaction are possible where all physical processes required for human life work in a similar way as in our universe (\citealp{Harnik:2006}, p. 1; \citealp{Grohs:2018}, p. 21). The possibility to remove the weak interaction from the standard model of physics to obtain a wide range of habitable universes means that our universe is not optimal with respect to the conciseness of its description. This is because a universe without the weak interaction has a simpler description than ours. Significantly enough, it has been argued that the vast majority of habitable universes do not have an observable weak interaction (\citealp{Gedalia:2011}, p. 4). Consequently, the maximization of the simplicity of a universe description does not seem to be a viable criterion to justify that our universe has a high likelihood. In other words, we may use the ontological Occam's razor to shave off our universe on the grounds that there are other universes equally hospitable to life, which are specified by simpler descriptions.

\section{Positivist universe selection\label{sec:Positivist-universe-selection}}

The number of possible states of the observable universe is finite (\citealp{Frampton:2009}, p. 1; \citealp{Egan:2010}, p. 1830). Let us imagine, to a first approximation, the possible histories of the universe as functions that give the current state of the universe for each Planck time since the Big Bang. The exact number of possible histories is not essential, provided that it is a finite number. The number of possible histories depends exponentially on the number of time instants $t$ and the number of possible states $s$, so that it is $O(s^t)$. The vast majority of such histories are uncompressible, i.e., their descriptions would not fit in the universe (\citealp{Shen:2017}, p. 230; \citealp{Li:2019}, p. 275). The positivist stance addresses this fact by focusing on counting the number of histories that comprise a configuration of the universe rather than trying to describe it concisely. Therefore, the positivist likelihood criterion assigns a probability $P(x)$ to a configuration of the universe $x$ as the quotient of the number of histories comprised by $x$, divided by the total number of possible histories.

Intelligent life could exist with less predictable physical laws. We could have many small departures from predictability, which would not be severe enough to threaten humankind so that civilization could have developed. Interestingly enough, we can not use the anthropic principle to discard all these alternative configurations of the universe, i.e., intelligent observers could live and thrive in those configurations. 
 
 Hence, from a positivist point of view, it is remarkably improbable that we observe a universe as predictable as ours. The argument for this goes as follows. Given any universe with any physical laws which allow for intelligent life, we can always introduce `perturbations' (unpredictable events) in many ways so that intelligent life is still possible. Each perturbed configuration will have the same probability as the unperturbed configuration so that the probability of living in the unperturbed configuration is tiny. Next, I give two examples.
 
 %Such perturbations are tiny when compared to the amount of quantum mechanical unpredictable events that happen everywhere at all times in universe. Therefore, these perturbations they add a negligible amount of Kolmogorov complexity to the description of the alternative universes. Hence such universes cannot be discarded on the grounds of a Kolmogorov complexity likelihood criterion.
 
For the first example, let us consider 10,000 stars of our galaxy (excluding our Sun). For reasons that will be evident later, we will assume that those stars can be easily observed by telescopes, but they can not be seen by the naked eye. An unexplainable and readily observable event would be that any of these stars ceased to produce light for, say, 1 minute. We could also assume that the gravitational properties of the star would not be changed so as not to alter the trajectories of the objects in the galaxy. It is clear that such an event would not influence life on Earth in any way. Nevertheless, this kind of event is not observed in our universe. How many of these events could happen in a year? We would have $2^{365 \cdot 24 \cdot 60 \cdot 10000}$ possible combinations of events, ranging from the case where no events happen, i.e., no star goes dark, to the case where all events happen, i.e., the 10,000 stars stay dark for all the year. All of these cases would be unexplainable by current physics except one: the combination where no `stellar blackout' occurs. None of them would affect life on Earth. If we assume that each of these combinations comprises the same number of universe histories, then all of them have the same probability. Consequently, we arrive at a probability of $2^{-365 \cdot 24 \cdot 60 \cdot 10000}$ of observing no stellar blackout.

The second example is based on microscopic events rather than stellar ones. There could be random changes in atoms, not explainable by physical laws, that would not affect life. It could be imagined that some atoms of elements irrelevant for biochemistry could transmute spontaneously and randomly into other biologically irrelevant elements without following the laws of physics. Here a low probability of unpredictable transmutation is assumed so that no cosmic chemistry processes are altered. For example, the half-life of some unstable atomic nuclei could vary slightly across time and space with respect to their standard values, which are regarded as invariant in time and space in our universe (\citealp{Pomme:2015}, p. 51). If the variation was small and the involved nuclei were not particularly relevant for cosmic or biological processes, then such unpredictable variation would not affect the possibility of intelligent life. Moreover, it would go unnoticed through human history until the half-life of atomic nuclei could be measured. The number of alternative universe histories that can be constructed in this way is much larger than those histories that exactly follow our laws of physics.

%The Kolmogorov complexity of such alternatives depend on how short the description of the unpredictable events can be. The position and time of the blackout events should be specified, and depending on the regularity of the pattern of the blackouts, this would lead to a different complexity. The more regular the pattern, the smaller Kolmogorov complexity.

It must be noted that the perturbed configurations that I am considering must have a small amount of unpredictability. There could be an entirely random universe with intelligent beings in it. However, that world would not be comprehensible for these intelligent beings. They would be much like Boltzmann brains. On the contrary, intelligence as we know it emerges from a moderately predictable environment, i.e., one where past experiences give us valuable information about the future. There can not be learning if the solutions to the problems found in the past will not work in the future. Intelligence does not evolve in a completely anarchic universe. Therefore, I restrict my attention to perturbed configurations that do not endanger the regularity of everyday events as perceived by intelligent beings.

Let us examine some possible objections to the examples:

\begin{enumerate}
\item It could be argued that the perturbations could endanger intelligent life. That is, if stellar blackouts were common, then the Sun should experience them too. Alternatively, the variations of the half-life of the atomic nuclei should also affect biochemically essential elements, thus preventing intelligent life. Nevertheless, this misses the point: we can choose any 10,000 stars to undergo blackouts while leaving the rest of the stars of the universe as they are, and we can choose any atomic nuclei that do not interfere with biology. This is because it is not necessary to provide a rational explanation about the selection of the stars or the nuclei, as long as the selection does not endanger intelligent life. Moreover, the selection does not need to follow any probability distribution nor exhibit any kind of regularity. Any selection will do since it will lead to roughly the same number of universe histories.

\item Another possible objection is that the human mind adapts to its environment across the ages so that anything we see as an irregularity would be seen as normal by the intelligent beings of these alternative universes. This can be readily refuted: in the previous examples, the human brain would have formed in the alternative stories exactly the same as in our universe, since the anomalous stars and nuclei could not be noticed until the invention of the telescope and atomic decay detectors, respectively. Hence, only at contemporary times, we would have discovered that the entire universe obeys the physical laws we know, except for those 10,000 stars, or that the half-life is constant across time and space except for some atomic nuclei. It is reasonable to think that such irregularities would have puzzled scientists, but they certainly would have allowed civilization to continue, while the organization of the human brain could not have been affected by this fact since telescopes and atomic decay detectors are very recent when compared to the time scale of the evolution of the human brain.

\item The irregularities might be seen as parts of the universe to be explained by more elaborate physical theories. This does not fly because the irregularities I am talking about can not be predicted by any law. In the examples, we could choose the stars or the kinds of atomic nuclei completely at random without following any pattern. 

\end{enumerate}

Consequently, there are many ways to obtain slightly perturbed versions of our own universe that contain anarchic aspects. The perturbed versions have a higher probability of occurring because they comprise a larger number of universe histories.

\section{Discussion}

In this work, two key assumptions have been made. The first one is that the object of our study is the observable universe, which is finite. Many approaches to cosmology assume that the universe is infinite or that there are infinite universes that comprise a multiverse (\citealp{Vilenkin:2006}, p. 204; \citealp{Susskind:2006}, chapter 13). Nevertheless, the physical existence of infinities has been termed as dubious or even unscientific because it is untestable (\citealp{Coley:2019}, p. 7; \citealp{Ellis:2014}, pp. 16-17). No matter whether they are real, physical infinities are not incorporated in our study because they would lead to infinite-length descriptions and infinite universe histories, thereby rendering the quantitative evaluation of the likelihood of the alternative universes unfeasible. The second assumption is that the values of physical constants and magnitudes have finite precision. Numeric values that must be specified with infinite precision lead to infinite descriptions, and furthermore, they would prevent the physical laws from producing effective predictions because the predictions would be uncomputable. As mentioned in Section \ref{sec:Preliminaries}, arbitrarily high precision numbers are still allowed since they do not prevent in principle that physical predictions are computed, provided that the required computational power is available.

Next, we discuss the implication of the possible existence of a Theory of Everything (ToE) in physics on the likelihood of a universe. In what follows, a ToE is understood as a set of laws for the entire universe that can be expressed in a short and simple description. A description of the laws of the universe must be short in order to qualify as a ToE. Otherwise, the description of the universe would be a plain enumeration of ground facts, as mentioned before (Section \ref{sec:Transcendentalist-universe-selection}). In other words, a ToE must by definition be simple, as highlighted in (\citealp{Hutter:2010}, p. 342). At the time of writing this work, no one knows whether a ToE exists (\citealp{Barnes:2012}, p. 562). If there is no ToE, then nothing changes with respect to the already presented arguments. If there is a ToE, then our argumentation for the positivist position (Section \ref{sec:Positivist-universe-selection}) remains unchanged since it does not make use of the laws of physics. The first proposed argument for the transcendentalist position (Section \ref{sec:Transcendentalist-universe-selection}) would still hold provided that the acknowledged fact that the randomness of quantum mechanics is irreducible (\citealp{Khrennikov:2019}, pp. 1-2; \citealp{Khrennikov:2016}, pp. 207-208; \citealp{Martinez:2018}, p. 2) is still valid in the ToE. The validity of the second argument that has been proposed for the transcendentalist view would depend on whether it is possible to remove the weak interaction from the ToE with an increase of the simplicity of the ToE. That is, if the ToE without the weak interaction was simpler than the ToE of our universe, then the second argument would still be valid.

It is also important to analyze the relation of our argument with the set of interpretations of quantum mechanics where the evolution of the universe is deterministic. In these interpretations, the wave function evolves deterministically, so there are no random events. For the sake of our argument, it does not matter whether the universe is deterministic. If the universe is deterministic, then the initial state of the universe must be specified in a very fine detail so that the combination of the initial state with the deterministic laws lead to the present state. Such specification of the initial state of the universe would require a very long specification, just like the description of the random events for a non-deterministic universe.

The significance of the presented arguments could be criticized on the grounds that it is not necessary that our universe is optimal with respect to the likelihood. It could be less likely than the most likely universe but still a probable one. In order to address this matter, the amount of alternative universes which are more likely than ours must be estimated. For each fundamental physical constant, each numeric initial condition, and each numeric specification of an unpredictable event of a universe, there is a countable infinity of possible values that it could have that are as close as desired to a reference value, and that can be represented finitely. In other words, for any given universe, there is a countable infinity of perturbed versions of that universe which are as close as desired to it. This holds for our universe and for any of the alternatives that we have presented before. Consequently, for all the examples that we have presented, there is a countable infinity of perturbed versions of each example that are more likely than our universe. Since the probabilities of the universes more likely than ours must add to a nonzero number, it follows that there must be universes in that set with arbitrarily small probabilities. Therefore, the probability of our universe, that is lower than the probability of each of those universes, must be exactly zero.

\section{Conclusions}

The evaluation of the likelihood of our universe with respect to other possibilities is a difficult task. There are no observations of such possibilities, so their characterization cannot be based on experimental data. The regularity of our universe, i.e., the fact that it obeys a set of physical laws, is seen as a remarkable feature that must be taken into account when assessing those likelihoods. It is also important to consider that, according to the mainstream view in current physics, there is irreducible unpredictability at the microscopic level. 

Two approaches to evaluating the likelihood of a possible universe have been identified, namely the transcendentalist and the positivist ones. Both are incommensurable because they advocate incompatible likelihood criteria. Consequently, they must be studied separately.

The transcendentalist strategy can be formalized by the criterion of simplicity of a description of a universe. This strategy implies that our universe is less simple than it should be to optimize the criterion due to the possibility of universes with pseudorandom quantum fluctuations and universes without weak force.

The positivist stance is formalized by the history counting criterion to assign likelihoods. Under this criterion, our universe is simpler than it should to be in order to be optimal since unpredictable perturbations can be introduced that do not affect intelligent life.

Therefore, neither the transcendentalist nor the positivist approaches to universe selection are able to properly justify a high likelihood of our own universe, as compared to other alternative universes.


\bibliographystyle{spbasic}      % basic style, author-year citations
\bibliography{main}   % name your BibTeX data base

\end{document}